\documentstyle[osa,manuscript]{revtex}

\begin{document}
\title{{\small {\it Foundations of Physics Letters }{\bf 12}, 251-254 (1998).
\bigskip \bigskip \bigskip } \\
{\normalsize {\bf ORTHOGONALITY AND BOUNDARY CONDITIONS IN QUANTUM MECHANICS
\bigskip}} }
\author{{\normalsize {\bf Alexander Gersten \bigskip }}}
\address{{\normalsize {\it Department of Physics }} \\
{\normalsize {\it Ben-Gurion University of the Negev \thanks{%
Permanent address.} }} \\
{\normalsize {\it Beer-Sheva 84105, Israel}} \\
{\normalsize and} \\
{\normalsize {\it Department of Physics}} \\
{\normalsize {\it City College of the City University of New York}} \\
{\normalsize {\it New York, New York 10031 \bigskip \bigskip}}}
\date{{\normalsize Received 27 December 1996 \bigskip \bigskip}}
\maketitle

{\normalsize \ One-dimensional particle states are constructed according to
orthogonality conditions, without requiring boundary conditions. Free
particle states are constructed using Dirac's delta function orthogonality
conditions. The states (doublets) depend on two quantum numbers: energy and
parity (''+'' or ''-''). With the aid of projection operators the particles
are confined to a constrained region, in a way similar to the action of an
infinite well potential. From the resulting overcomplete basis only the
mutually orthogonal states are selected. Four solutions are found,
corresponding to different non-commuting Hamiltonians. Their energy
eigenstates are labeled with the main quantum number n and parity ''+'' or
''-''. The energy eigenvalues are functions of n only. The four cases
correspond to different boundary conditions: (I) the wave function vanishes
on the boundary (energy levels: 1}$^{+}${\normalsize ,2}$^{-}${\normalsize ,3%
}$^{+}${\normalsize ,4}$^{-}${\normalsize ,...), (II) the derivative of the
wavefunction vanishes on the boundary (energy levels 0}$^{+}${\normalsize ,1}%
$^{-}${\normalsize ,2}$^{+}${\normalsize ,3}$^{-}${\normalsize ,...), (III)
periodic boundary conditions (energy levels: 0}$^{+}${\normalsize ,2}$^{+}$%
{\normalsize ,2}$^{-}${\normalsize ,4}$^{+}${\normalsize ,4}$^{-}$%
{\normalsize 6}$^{+}${\normalsize ,6}$^{-}${\normalsize ,...), (IV) periodic
boundary conditions (energy levels: 1}$^{+}${\normalsize ,1}$^{-}$%
{\normalsize ,3}$^{+}${\normalsize ,3}$^{-}${\normalsize ,5}$^{+}$%
{\normalsize ,5}$^{-}${\normalsize ,...). Among the four cases, only
solution (III) forms a complete basis in the sense that any function in the
constrained region, can be expanded with it. By extending the boundaries of
the constrained region to infinity, only solution (III) converges uniformly
to the free particle states. Orthogonality seems to be a more basic
requirement than boundary conditions. By using projection operators,
confinement of the particle to a definite region can be achieved in a
conceptually simple and unambiguous way, and physical operators can be
written so that they act only in the confined region. \bigskip \bigskip }

{\normalsize \raggedright Key words: confinement, orthogonality, boundary
conditions, Dirac's formalism, projection operators.}

\section{INTRODUCTION}

In quantum mechanics the physical states corresponding to different
eigenvalues are orthogonal [1-3]. Therefore orthogonality is much more than
a convenient tool for expanding functions in terms of a suitable orthogonal
basis. On the other hand the space of quantum states is a normed space and
normalization of states is required. Usually, when the coordinates are
extending to infinity, in order to secure normalizability, boundary
conditions have to be imposed. But this procedure in quantum mechanics
becomes too rigid. We show in the case of an infinite square well (Sec. 3),
for which normalizability is always secured, that by imposing boundary
conditions, part of the solutions can be missing. In this case, in order to
get more solutions, we impose orthogonality alone. In this way new solutions
are obtained. The energy levels are not changed. The difference is that with
the boundary conditions the energy states are singlets, with the
orthogonality conditions the energy states may be also doublets. This brings
also changes in the bases of the vector space. This is extensively discussed
in Sec. 3.

In Sec. 2 we obtain the doublet energy states of a free particle by imposing
the Dirac's delta function conditions of orthogonality only, without
boundary conditions. Thus working with Dirac's representation theory of
quantum mechanics (1), one can get a larger basis of states and new
possibilities compared to quantum mechanics which is defined in Hilbert
space. Working in the Dirac's abstract space of vector states does not
require boundary conditions [4], as long as they are not projected onto the
coordinate states.

In Sec. 3 we find solutions for the infinite square well problem by
redefining it so that by using a proper projection operator, the free
particle solution is projected into the confined subspace inside the
infinite square well. The projected free particle solution forms an
overcomplete nonorthogonal basis. We find solutions to the problem after
sifting from the overcomplete solutions the orthogonal ones. In this way we
find additional solutions to the ones usually quoted in textbooks [2,3]. The
new solutions satisfy new boundary conditions. But the solutions so obtained
are not resulting only from the possibility to reformulate boundary
conditions. We prove that there are no solutions to mixed boundary
conditions, namely vanishing of the wavefunction at one end and vanishing of
the derivative at the other end and vice versa.

\bigskip

\section{ENERGY EIGENVECTORS OF A FREE PARTICLE}

In this section we derive the energy eigenvectors of a free particle with
mass m and momentum p, in one dimension, on the infinite x axis. We shall do
it first in an abstract way, without projecting them into coordinates. The
free particle energy (Hamiltonian) operator is

$$
\widehat{H}=\frac{\widehat{p}^{2}}{2m},\eqno(2.1) 
$$
with the energy eigenvalue E satisfying the equation

$$
\widehat{H}|E\rangle =\frac{\widehat{p}^{2}}{2m}|E\rangle =E|E\rangle ,\eqno%
(2.1a) 
$$
where $\widehat{p}$ is the momentum operator and p its eigenvalue satisfying
the eigenvalue equation 
$$
\widehat{p}|p\rangle =p|p\rangle .\eqno(2.2) 
$$
\ \ \ \ 

The energy eigenvectors are obviously also the eigenvectors of \ $\widehat{p}%
^{2}$. We shall find them in the following way. Let us first note that 
$$
\widehat{p}^{2}|p\rangle =p^{2}|p\rangle ;\ \ \ \ \widehat{p}^{2}|-p\rangle
=p^{2}|-p\rangle ;\eqno(2.3) 
$$
therefore any combination of these two states will be also an eigenvector of
the momentum squared operator: 
$$
|p^{2}\rangle =A|p\rangle +B|-p\rangle ,\eqno(2.4) 
$$
where A and B are constants. Following Dirac [1], we shall require the
following normalization: 
$$
\langle p|p^{\prime }\rangle =\delta \left( p-p^{\prime }\right) ,\eqno(2.5) 
$$
$$
\langle p^{2}|p^{\prime 2}\rangle =\delta \left( p^{2}-p^{\prime 2}\right) =%
\frac{1}{2|p|}\left[ \delta \left( p-p^{\prime }\right) +\delta \left(
p+p^{\prime }\right) \right] .\eqno(2.6) 
$$
One can prove that the only combination of (2.4) that satisfies Eq. (2.6)
(up to a phase factor) is 
$$
|p_{+}^{2}\rangle =\frac{1}{2\sqrt{|p|}}\left( |p\rangle +|-p\rangle \right)
.\eqno(2.7) 
$$
We can find a second possibility (again up to a phase factor) in the form 
$$
|p_{-}^{2}\rangle =\frac{{\normalsize sign}(p)}{2i\sqrt{|p|}}\left( {\large |%
}p{\large \rangle }-{\large |}-p{\large \rangle }\right) =\frac{1}{2i\sqrt{%
|p|}}\left( {\Large |}|p|{\large \rangle }-{\Large |}-|p|{\large \rangle }%
\right) ,\eqno(2.8) 
$$
where 
$$
{\normalsize sign}(p)=\left\{ 
\begin{array}{c}
\ \ 1,\ \ \ \ \ for\ \ p>0, \\ 
\ \ 0,\ \ \ \ \ for\ \ p=0, \\ 
-1,\ \ \ \ \ for\ \ p<0.
\end{array}
\right. \eqno(2.9) 
$$

One can easily verify that the vectors (2.7) and (2.8) are mutually
orthogonal. Moreover, they form a complete orthonormal basis, which can be
proved in the following way: 
$$
\textstyle\int%
\limits_{0}^{\infty }\left( |p_{+}^{2}\rangle \langle
p_{+}^{2}|+|p_{-}^{2}\rangle \langle p_{-}^{2}|\right) dp^{2}=%
\textstyle\int%
\limits_{0}^{\infty }\left( |p\rangle \langle p|+|-p\rangle \langle
-p|\right) dp=%
\textstyle\int%
\limits_{-\infty }^{\infty }|p\rangle \langle p|dp=\widehat{I,}\eqno(2.10) 
$$
where $\widehat{I}$ is the identity operator. The above results allow us to
conclude that for each energy eigenvalue there are two degenerate states
having the same energy eigenvalue $E_{+}=E_{-}$ for which the following
relations are satisfied: 
$$
E_{+}=E_{-}=\frac{p^{2}}{2m}\equiv E,\ \ \ \ \langle E_{+}|E_{-}\rangle =0,%
\eqno
$$
$$
\ \langle E_{+}|E_{+}^{^{\prime }}\rangle =\langle E_{-}|E_{-}^{^{\prime
}}\rangle =\delta \left( E-E^{\prime }\right) =\delta \left( \frac{p^{2}}{2m}%
-\frac{p^{\prime 2}}{2m}\right) =\frac{m}{|p|}\left[ \delta \left(
p-p^{\prime }\right) +\delta \left( p+p^{\prime }\right) \right] ,\eqno%
(2.11) 
$$
$$
|E+\rangle =\sqrt{\frac{m}{2|p|}}\left( {\large |}|p|{\large \rangle }+%
{\large |}-|p|{\large \rangle }\right) ,\ \ \ \ |E_{-}\rangle =-i\sqrt{\frac{%
m}{2|p|}}\left( {\large |}|p|{\large \rangle }-{\large |}-|p|{\large \rangle 
}\right) .\eqno(2.12) 
$$
The projection into x space proceeds via the momentum wavefunctions

$$
\langle x|p\rangle =\frac{1}{\sqrt{2\pi \hbar }}\exp \left( \frac{ipx}{\hbar 
}\right) ,\eqno(2.13) 
$$
(normalized according to (2.5)) and leads to twofold energy degenerate
wavefunctions 
$$
\langle x|E_{+}\rangle =\sqrt{\frac{m}{\pi \hbar |p|}}\cos \left( \frac{|p|x%
}{\hbar }\right) ,\eqno(2.14a) 
$$
$$
\langle x|E_{-}\rangle =\sqrt{\frac{m}{\pi \hbar |p|}}\sin \left( \frac{|p|x%
}{\hbar }\right) ,\eqno(2.14b) 
$$
According to their transformation properties with respect to x we will call
the solution (2.14a) a positive parity wavefunction and the solution (2.14b)
the negative parity wavefunction. Accordingly we can define the energy
states with a definite parity:

$$
|E,\Pi \rangle =\left\{ 
\begin{array}{c}
|E_{+}\rangle ,\ \ \ \ for\ \ \Pi =1, \\ 
|E_{-}\rangle ,\ \ \ \ \ for\ \Pi =-1,
\end{array}
\right. \eqno(2.15) 
$$
and write the completeness relation in the form 
$$
\widehat{I}=%
\textstyle\int%
\limits_{0}^{\infty }\left( |E,1\rangle \langle E,1|+|E,-1\rangle \langle
E,-1|\right) dE.\eqno(2.16) 
$$

\section{THE INFINITE POTENTIAL WELL}

Let us consider the case of a particle in an infinite potential well. The
potential is defined as:

$$
V(x)=\left\{ 
\begin{array}{c}
0,\ \ \ -a\leq x\leq a, \\ 
\infty ,\ \ \ \ \ \ \ \ \ |x|>a,
\end{array}
\right. \eqno(3.1) 
$$
i.e. the particle is confined in a ''bandwidth'' of width $a$ [4]. The time
independent Schroedinger equation (in the stationary state) has the form 
$$
\frac{-\hbar ^{2}}{2m}\frac{\partial ^{2}}{\partial x^{2}}\langle x|E\rangle
=E\langle x|E\rangle ,\ \ \ \ \langle a|E\rangle =\langle -a|E\rangle =0,%
\eqno(3.2) 
$$
where $\hbar $ is the Planck constant, m the mass of the particle, and E the
energy eigenvalue.

By using orthogonality considerations rather than boundary conditions we
will find additional solutions, such that the energy states will be twofold
degenerate. Let $\widehat{H}$ be the Hamiltonian (the energy operator). As
an observable it is required to be Hermitian; therefore $\langle E|\widehat{H%
}|E^{\prime }\rangle =E$ $\langle E|E^{\prime }\rangle =E^{\prime }\langle
E|E^{\prime }\rangle ;$ and after subtracting both sides, we find that
physical states have to satisfy the orthogonality condition 
$$
\left( E-E^{\prime }\right) ^{\prime }\langle E|E^{\prime }\rangle =0.\eqno%
(3.3) 
$$
These conditions are quite general and do not explicitly have boundary
conditions. Boundary conditions are required when the scalar product has to
be finite. But in the case of an infinite square well there is no problem
that the scalar products and normalizations will be finite; therefore the
condition (3.3) seems to be more adequate than boundary conditions.

The eigenvalues and orthogonal eigensolutions of Eq. (3.2), as they usually
appear in textbooks [2,3], are 
$$
E_{n}=\frac{p_{n}^{2}}{2m}=\frac{\hbar ^{2}k_{n}^{2}}{2m},\ \ \ \ \ k_{n}=%
\frac{n\pi }{2a},\ \ \ \ \ n=1,2,3,...,\eqno(3.4) 
$$
$$
\langle x|E_{n-}\rangle =\frac{1}{\sqrt{a}}\sin \left( k_{n}x\right) ,\ \ \
\ for\ \ n=2,4,6,...,\eqno(3.5a) 
$$
$$
\langle x|E_{n+}\rangle =\frac{1}{\sqrt{a}}\cos \left( k_{n}x\right) ,\ \ \
\ for\ \ n=1,3,5,...,\eqno(3.5b) 
$$

Let us use a different approach for finding the energy solutions, namely
that of Eq. (3.3). In Sec. 2 we found twofold degenerate energy solutions
for the free particle case:

$$
\langle x|E_{+}\rangle =\sqrt{\frac{m}{\pi \hbar |p|}}\cos \left( \frac{|p|x%
}{\hbar }\right) ,\ \ \ \ \langle x|E_{-}\rangle =\sqrt{\frac{m}{\pi \hbar
|p|}}\sin \left( \frac{|p|x}{\hbar }\right) .\eqno(3.6) 
$$
Let us project the whole free particle vector space into a subspace given by
the projection operator: 
$$
\widehat{X}_{a}=%
\textstyle\int%
\limits_{-a}^{a}|x\rangle \langle x|dx,\ \eqno(3.7) 
$$
i.e. we now confine the free particle to exist only in $-a\leq x\leq a$. In
section 2 the normalizations of the free particle states were 
$$
E_{+}=E_{-}=\frac{p^{2}}{2m}=E,\ \ \ \ \langle E_{+}|E_{-}^{\prime }\rangle
=0,\eqno
$$
$$
\ \langle E_{+}|E_{+}^{\prime }\rangle =\ \langle E_{-}|E_{-}^{\prime
}\rangle =\delta \left( E-E^{\prime }\right) =\delta \left( \frac{p^{2}}{2m}-%
\frac{p^{\prime 2}}{2m}\right) =\frac{m}{\sqrt{|pp^{\prime }|}}\left[ \delta
\left( p-p^{\prime }\right) +\delta \left( p+p^{\prime }\right) \right] .%
\eqno(3.8) 
$$
Now, in the constrained vector subspace, the delta functions will be
replaced by (the incomplete delta functions [5] ): 
$$
\langle E_{+}|\widehat{X}_{a}|E_{+}^{\prime }\rangle =\langle E_{-}|\widehat{%
X}_{a}|E_{-}^{\prime }\rangle =\frac{m}{\sqrt{|pp^{\prime }|}}\left( \frac{%
\sin \left[ a\left( p-p^{\prime }\right) /\hbar \right] }{\pi \left(
p-p^{\prime }\right) }+\frac{\sin \left[ a\left( p+p^{\prime }\right) /\hbar %
\right] }{\pi \left( p+p^{\prime }\right) }\right) .\eqno(3.9) 
$$
For $p=p^{\prime },$\ we have 
$$
\langle E_{+}|\widehat{X}_{a}|E_{+}\rangle =\langle E_{-}|\widehat{X}%
_{a}|E_{-}\rangle =\frac{ma}{\pi \hbar |p|},\eqno(3.9a) 
$$
According to Eq. (3.5), physical states, which are eigenstates of Hermitian
operators (observables), should be orthogonal for different energies. In Eq.
(3.9) this will happen if 
$$
p-p^{\prime }=\frac{n_{1}\pi \hbar }{a},\ \ \ \ \ and\ \ p+p^{\prime }=\frac{%
n_{2}\pi \hbar }{a},\ \ \ \ \ n_{1}\ and\ n_{2}\ integers\eqno(3.10) 
$$
There are two solutions of Eqs. (3.10) (characterized by the positive
integer n): 
$$
(I):\ \ \ p_{2n}=\frac{n\pi \hbar }{a},\eqno(3.11a) 
$$
$$
(II):\ \ p_{2n+1}=\frac{n\pi \hbar }{a}+\frac{\pi \hbar }{2a}.\eqno(3.11b) 
$$
Let us describe the wavefunctions, the solutions to (3.1), in the following
way: 
$$
\langle x|E\left( |p|\right) ,\Pi \rangle =\left\{ 
\begin{array}{c}
\frac{1}{\sqrt{a}}\cos {\normalsize (}\frac{|p|x}{\hbar }{\normalsize ),}\ \
\ \ \ for\ \Pi =1, \\ 
\frac{1}{\sqrt{a}}\sin {\normalsize (}\frac{|p|x}{\hbar }{\normalsize ),}\ \
\ \ \ for\ \Pi =-1,
\end{array}
\right. \eqno(3.12) 
$$
where $\Pi $ is the parity eigenvalue. One should note that the solutions
(3.11a) and (3.11b) are valid for both parities (see Eq. (3.8)). Therefore
we can get four different orthonormal solutions: 
$$
(I):\left. 
\begin{array}{c}
\langle x|E\left( p_{2n-1}\right) ,1\rangle =\frac{1}{\sqrt{a}}\cos \left( 
\frac{(2n-1)\pi x}{2a}\right) , \\ 
\langle x|E\left( p_{2n}\right) ,-1\rangle =\frac{1}{\sqrt{a}}\sin \left( 
\frac{n\pi x}{a}\right) ;\ \ \ \ \ \ \ \ \ \ 
\end{array}
\right. \eqno(3.13a) 
$$

$$
(II):\left. 
\begin{array}{c}
\langle x|E\left( p_{2n}\right) ,1\rangle =\frac{1}{\sqrt{a}}\cos \left( 
\frac{n\pi x}{a}\right) ,\ \ \ \ \ \ \ \ \ \ \ \  \\ 
\langle x|E\left( p_{2n-1}\right) ,-1\rangle =\frac{1}{\sqrt{a}}\sin \left( 
\frac{(2n-1)\pi x}{2a}\right) ;
\end{array}
\right. \eqno(3.13b) 
$$

$$
(III):\ \ \ \langle x|E\left( p_{2n}\right) ,1\rangle ,\ \ \ \ \langle
x|E\left( p_{2n}\right) ,-1\rangle ,\ \ \ \ \ \ \ \eqno(3.13c) 
$$

$$
(IV):\ \ \ \langle x|E\left( p_{2n-1}\right) ,1\rangle ,\ \ \ \ \langle
x|E\left( p_{2n-1}\right) ,-1\rangle ,\ \ \ \ \ \ \eqno(3.13d) 
$$
where solution (I) corresponds to the standard solution (3.6) for which the
wavefunctions vanish on the boundary. Solution (II) corresponds to boundary
conditions for which the derivatives of the wavefunctions vanish on the
boundary. Solution (III) is the (complete) basis of Fourier series, with
periodic (symmetric) boundary conditions. Solution (IV) corresponds to
periodic (antisymmetric) boundary conditions. The sum of solutions (III) and
(IV) is equal to the sum of solutions (I) and (II); therefore only three
solutions are linearly independent. The main problem with solutions (I-IV)
is that although each solution separately is an orthonormal basis, the
solutions are not always orthogonal to each other. This means that they must
correspond to different Hamiltonians. Therefore the projection into a
confined subspace generated four Hamiltonians (three linearly independent)
for which physically acceptable solutions represent particles confined in
the coordinate subspace. We shall discuss in more detail these Hamiltonians
in Sec. 4.

Before that, let us consider mixed boundary conditions, i.e. vanishing of
the wavefunction at x=a and vanishing of the derivative at x=-a or vice
versa. We will prove by contradiction that there are no solutions for these
problems. Let us first assume that there are solutions. Let us call such a
solution $\psi $(x), which satisfies 
$$
\psi (a)=\psi ^{\prime }(-a)=0,\eqno(3.14) 
$$
The most general form of the solution is 
$$
\psi (x)=A\cos (hx)+B\sin (hx),\eqno(3.15) 
$$
where $A,B$ and $h$ are constants. After substituting the boundery
conditions (3.14) into Eq. (3.15) we obtain 
$$
A\cos (ha)+B\sin (ha)=0,\eqno(3.16a) 
$$
$$
hA\sin (ha)+hB\cos (ha)=0.\eqno(3.16b) 
$$
Eqs. (3.16a) and (3.16b) can be rewritten as: 
$$
A\cos (ha)=-B\sin (ha),\eqno(3.17a) 
$$
$$
A\sin (ha)=-B\cos (ha).\eqno(3.17b) 
$$
Dividing Eq. (3.17a) by Eq. (3.17b), we obtain: 
$$
\frac{\cos (ha)}{\sin (ha)}=-\frac{\sin (ha)}{\cos (ha)},\eqno(3.18) 
$$
which can be rewritten as 
$$
\cos ^{2}(ha)+\sin ^{2}(ha)=0,\eqno(3.19) 
$$
which is a contradiction as the right hand-side of Eq. (3.19) should be
equal to one. In this way we arrive at twofold degenerate energy states. We
see that in the case of the infinite well the orthogonality requirement
leads to all solutions, while by imposing boundary conditions on the
wavefunction some of the solutions are missing. In the next section we
demonstrate that we could have guessed this result by considering the
completeness relation in the x subspace constrained by the well.

\section{THE HAMILTONIANS AND COMPLETENESS OF THE BASES}

The four Hamiltonians, corresponding to the solutions (I-IV), can be
represented in the following way:

$$
\begin{array}{c}
H_{I}=%
\mathop{\textstyle\sum}%
\limits_{n=1}^{\infty }E\left( p_{2n-1}\right) \widehat{X}_{a}{\large |}%
E\left( p_{2n-1}\right) ,1{\large \rangle \langle }E\left( p_{2n-1}\right) ,1%
{\large |}\widehat{X}_{a} \\ 
\ \ \ \ \ \ \ \ \ \ \ \ \ \ \ \ \ \ \ +%
\mathop{\textstyle\sum}%
\limits_{n=1}^{\infty }E\left( p_{2n}\right) \widehat{X}_{a}{\large |}%
E\left( p_{2n}\right) ,-1{\large \rangle \langle }E\left( p_{2n}\right) ,-1%
{\large |}\widehat{X}_{a},
\end{array}
\eqno(4.1) 
$$

$$
\begin{array}{c}
H_{II}=%
\mathop{\textstyle\sum}%
\limits_{n=1}^{\infty }E\left( p_{2n-1}\right) \widehat{X}_{a}{\large |}%
E\left( p_{2n-1}\right) ,-1{\large \rangle \langle }E\left( p_{2n-1}\right)
,-1{\large |}\widehat{X}_{a} \\ 
\ \ \ \ \ \ \ \ \ \ \ \ \ \ \ \ \ \ \ +%
\mathop{\textstyle\sum}%
\limits_{n=1}^{\infty }E\left( p_{2n}\right) \widehat{X}_{a}{\large |}%
E\left( p_{2n}\right) ,1{\large \rangle \langle }E\left( p_{2n}\right) ,1%
{\large |}\widehat{X}_{a},
\end{array}
\eqno(4.2) 
$$
$$
\begin{array}{c}
H_{III}=%
\mathop{\textstyle\sum}%
\limits_{n=1}^{\infty }E\left( p_{2n}\right) \widehat{X}_{a}{\large |}%
E\left( p_{2n}\right) ,-1{\large \rangle \langle }E\left( p_{2n-1}\right) ,-1%
{\large |}\widehat{X}_{a} \\ 
\ \ \ \ \ \ \ \ \ \ \ \ \ \ \ \ \ \ \ +%
\mathop{\textstyle\sum}%
\limits_{n=1}^{\infty }E\left( p_{2n}\right) \widehat{X}_{a}{\large |}%
E\left( p_{2n}\right) ,1{\large \rangle \langle }E\left( p_{2n}\right) ,1%
{\large |}\widehat{X}_{a},
\end{array}
\eqno(4.3) 
$$
$$
\begin{array}{c}
H_{IV}=%
\mathop{\textstyle\sum}%
\limits_{n=1}^{\infty }E\left( p_{2n-1}\right) \widehat{X}_{a}{\large |}%
E\left( p_{2n-1}\right) ,-1{\large \rangle \langle }E\left( p_{2n-1}\right)
,-1{\large |}\widehat{X}_{a} \\ 
\ \ \ \ \ \ \ \ \ \ \ \ \ \ \ \ \ \ \ +%
\mathop{\textstyle\sum}%
\limits_{n=1}^{\infty }E\left( p_{2n-1}\right) \widehat{X}_{a}{\large |}%
E\left( p_{2n-1}\right) ,1{\large \rangle \langle }E\left( p_{2n-1}\right) ,1%
{\large |}\widehat{X}_{a},
\end{array}
\eqno(4.4) 
$$
where $\widehat{X}_{a}$ is the projection (into the well) operator, given by
Eq. (3.7), and 
$$
E\left( p_{j}\right) =\frac{p_{j}^{2}}{2m},\ \ \ \ \ p_{j}=\frac{j\pi \hbar 
}{2a},\ \ \ \ \ j=0,1,2,3,...\ ,\eqno(4.5) 
$$
$$
|E\left( p_{j}\right) ,1\rangle =\frac{|p_{j}\rangle +|-p_{j}\rangle }{\sqrt{%
2}},\ \ \ \ \ |E\left( p_{j}\right) ,-1\rangle =\frac{|p_{j}\rangle
-|-p_{j}\rangle }{\sqrt{2}i},\ \ \eqno(4.6)\ 
$$
$$
\langle x|\widehat{X}_{a}|p_{j}\rangle =\frac{1}{\sqrt{2a}}\exp \left(
ip_{j}x\right) ,\ \ \ \ \ \langle p_{j}|p_{j}\rangle =1.\eqno(4.7) 
$$
The normalizations were chosen to agree with the normalizations of Eqs.
(3.13a-3.13d). The Hamiltonians (4.1-4.4) are linearly dependent: 
$$
H_{I}+H_{II}=H_{III}+H_{IV},\eqno(4.8) 
$$
and each one of them, separately, represents a particle confined to a well.
For instance, in the example 
$$
\begin{array}{c}
H_{I}\left( \widehat{X}_{a}|E\left( p_{2n-1}\right) ,1\rangle \right)
=E\left( p_{2n-1}\right) \left( \widehat{X}_{a}|E\left( p_{2n-1}\right)
,1\rangle \right) \\ 
H_{I}\left( \widehat{X}_{a}|E\left( p_{2n}\right) ,-1\rangle \right)
=E\left( p_{2n}\right) \left( \widehat{X}_{a}|E\left( p_{2n}\right)
,-1\rangle \right)
\end{array}
\eqno(4.9) 
$$
the states of Eq. (3.13a) (which vanish at the ends of the well) are
obtained. Note that the presence of the projection operator assures that no
particle can be found outside the well. Let us now consider the problem of
completeness of the bases (3.13a-3.13d). First we should specify what is the
meaning of completeness in the framework of the Dirac formalism. If we start
with the unconstrained vector space, then the completeness is expressed
through the identity operator $\widehat{I}$: 
$$
\widehat{I}=%
\textstyle\int%
\limits_{0}^{\infty }\left( |E,1\rangle \langle E,1|+|E,-1\rangle \langle
E,-1|\right) dE=%
\textstyle\int%
\limits_{-\infty }^{\infty }|p\rangle \langle p|dp=%
\textstyle\int%
\limits_{-\infty }^{\infty }|x\rangle \langle x|dx.\eqno(4.10) 
$$
If we constrain the coordinate space with the projection operator (3.7), all
states, including any basis states, has to be projected into the subspace.
For example the momentum basis has to be projected onto 
$$
|p\rangle \Longrightarrow \widehat{X}_{a}|p\rangle ,\eqno(4.11) 
$$
and completeness means: 
$$
\textstyle\int%
\limits_{-\infty }^{\infty }\widehat{X}_{a}|p\rangle \langle p|\widehat{X}%
_{a}dp=\widehat{X}_{a}\widehat{I}\widehat{X}_{a}=\widehat{X}_{a},\eqno(4.12) 
$$
i.e., in the subspace the projection operator becomes the identity operator
(for the subspace). The new basis (4.11) is overcomplete and nonorthogonal.
In the quantization process the eigenstates of the Hamiltonian can form an
orthogonal basis. Are the bases (3.13a-3.13d) complete? In order to answer
this question, let us construct the projection operators generating the
bases (3.13a-3.13d): 
$$
P_{I}=%
\mathop{\textstyle\sum}%
\limits_{n=1}^{\infty }\widehat{X}_{a}\left( {\large |}E\left(
p_{2n-1}\right) ,1{\large \rangle \langle }E\left( p_{2n-1}\right) ,1{\large %
|}+{\large |}E\left( p_{2n}\right) ,-1{\large \rangle \langle }E\left(
p_{2n}\right) ,-1|\right) \widehat{X}_{a},\eqno(4.13a) 
$$
$$
P_{II}=%
\mathop{\textstyle\sum}%
\limits_{n=1}^{\infty }\widehat{X}_{a}\left( {\large |}E\left(
p_{2n-1}\right) ,-1{\large \rangle \langle }E\left( p_{2n-1}\right) ,-%
{\normalsize 1}{\large |}+{\large |}E\left( p_{2n}\right) ,1{\large \rangle
\langle }E\left( p_{2n}\right) ,1{\large |}\right) \widehat{X}_{a},\eqno%
(4.13b) 
$$
$$
P_{III}=%
\mathop{\textstyle\sum}%
\limits_{n=1}^{\infty }\widehat{X}_{a}\left( {\large |}E\left( p_{2n}\right)
,-1{\large \rangle \langle }E\left( p_{2n-1}\right) ,-1{\large |}+{\large |}%
E\left( p_{2n}\right) ,1{\large \rangle \langle }E\left( p_{2n}\right) ,1%
{\large |}\right) \widehat{X}_{a},\eqno(4.13c) 
$$
$$
P_{IV}=%
\mathop{\textstyle\sum}%
\limits_{n=1}^{\infty }\widehat{X}_{a}\left( {\large |}E\left(
p_{2n-1}\right) ,-1{\large \rangle \langle }E\left( p_{2n-1}\right) ,-1%
{\large |}+{\large |}E\left( p_{2n-1}\right) ,1{\large \rangle \langle }%
E\left( p_{2n-1}\right) ,1{\large |}\right) \widehat{X}_{a},\eqno(4.13d) 
$$
with the linear dependence: 
$$
P_{I}+P_{II}=P_{III}+P_{IV}.\eqno(4.13e) 
$$
Is the projection operator $P_{I}$ equal to $\widehat{X}_{a}$? The answer is
negative. If we apply the above projection operators to an arbitrary
function f(x) we obtain functions with support in the subspace only, with
the properties 
$$
\langle x|P_{I}|f\rangle \Longrightarrow \langle \pm a|P_{I}|f\rangle =0,%
\eqno(4.14) 
$$
$$
\langle x|\widehat{X}_{a}|f\rangle \Longrightarrow
\lim\limits_{x\longrightarrow a}\langle \pm x|\widehat{X}_{a}|f\rangle
=\lim\limits_{x\longrightarrow a}f\left( \pm x\right) ,\eqno(4.15) 
$$
i.e., $P_{I}$ projects only to functions which are vanishing at the
boundary; therefore the basis (3.13a) can be complete only with respect to
functions vanishing on the boundary. The basis (3.13b) can be complete only
with respect to functions whose derivative vanishes on the boundary. The
basis (3.13c) is the basis for Fourier series expansion and is known to be
complete in the subspace. The basis (3.13d) is related to the basis (3.13c)
via 
$$
\langle x|E\left( p_{2n-1}\right) ,1\rangle =\cos \left( \frac{\pi x}{2a}%
\right) \langle x|E\left( p_{2n}\right) ,1\rangle +\sin \left( \frac{\pi x}{%
2a}\right) \langle x|E\left( p_{2n}\right) ,-1\rangle ,\eqno(4.16) 
$$
$$
\langle x|E\left( p_{2n-1}\right) ,-1\rangle =\cos \left( \frac{\pi x}{2a}%
\right) \langle x|E\left( p_{2n}\right) ,-1\rangle -\sin \left( \frac{\pi x}{%
2a}\right) \langle x|E\left( p_{2n}\right) ,1\rangle ,\eqno(4.17) 
$$
but any expansion in terms of the basis (3.13d) will miss a constant term
(which is present in the Fourier series), and for this reason the basis
(3.13d) is not complete. Up to that constant term, any expansion with the
basis (3.13d) can be brought to a non vanishing combination of Fourier
series; therefore it is complete for functions not having a constant term in
their Fourier expansion.

\section{SUMMARY AND CONCLUSIONS}

In this work one dimensional particle states were constructed according to
orthogonality conditions, without requiring boundary conditions. Free
particle states were constructed using Dirac's delta function orthogonality
conditions Eq. (2.11). The states, given in Eq. (2.12), depended on two
quantum numbers: energy and parity (''+'' or ''-''). The eigenvalues
depended only on the energy, therefore the states were doublets in energy
(with the exception of a singlet at zero energy). With the aid of the
projection operator of Eq. (3.7), the particles were confined to a
constrained region, in a way similar to the action of an infinite-well
potential. From the resulting overcomplete basis, only the mutually
orthogonal states are selected from Eq. (3.9). Four solutions were found:
Eqs. (3.13a-3.13d) , corresponding to different non-commuting Hamiltonians
of Eqs. (4.1-4.4). Their energy eigenstates were labeled with the main
quantum number n and parity ''+'' or ''-''. The energy eigenvalues were
functions of n only. The four cases corresponded to different boundary
conditions:

(I) The wave function vanishes on the boundary (energy levels: $%
1^{+},2^{-},3^{+},4^{-},...$). This is the standard model of the infinite
square well.

(II) The derivative of the wavefunction vanishes on the boundary (energy
levels $0^{+},1^{-},2^{+},3^{-},...$).

(II) Periodic boundary conditions (energy levels: $%
0^{+},2^{+},2^{-},4^{+},4^{-},6^{+},6^{-},...$). The eigenfunctions of Eq.
(3.13.c) coincide with the basis of the Fourier series.

(IV) Periodic boundary conditions (energy levels: $%
1^{+},1^{-},3^{+},3^{-},5^{+},5^{-},...$).

Among the four cases, only solution (III) forms a complete basis, in the
sense that any function in the constrained region, can be expanded with it.
By extending the boundaries of the constrained region to infinity, it seems
that only the doublets of solution (III) converge to the free particle
doublet states.

By confining a particle to a constrained subspace and requiring, as the
basic physical condition, the orthogonality of states, we obtained four
exclusive solutions, corresponding to different boundary conditions.
Therefore orthogonality seems to be a more basic requirement than boundary
conditions. By using projection operators, confinement of the particle to a
definite region can be achieved in a simple and unambiguous way. All
physical operators can be written so that they act only in the confined
region. This method seems to be superior to the boundary condition models.
\bigskip

\subsection*{ACKNOWLEDGMENTS.}

It is a pleasure to thank Gideon Erez, Joel Gersten, and Joseph Malinsky for
valuable discussions and encouragement.

\section*{REFERENCES}


\begin{enumerate}
\item  P.A.M. Dirac: {\it The Principles of Quantum Mechanics}, 4th edn.
(Clarendon, Oxford, 1958).

\item  L.I. Schiff: {\it Quantum Mechanics}, 3rd edn. (McGraw-Hill, New
York, 1968)

\item  L.D. Landau and E.M. Lifshitz: {\it Quantum Mechanics-
Nonrelativistic Theory}, 3rd edn. (Pergamon, New York, 1975)

\item  A. Gersten: {\it Ann. Phys.} 262, 47-72, 1998; 262, 73-104, 1998

\item  A. Gersten: {\it Found. Phys. Lett.} 11, 165-178, 1998
\end{enumerate}


\end{document}